\documentclass[preprint,showpacs,preprintnumbers,amsmath,amssymb,aps,pre]{revtex4}
\usepackage{graphicx}% Include figure files
\usepackage{dcolumn}% Align table columns on decimal point
\usepackage{bm}% bold math
\begin{document}
%\preprint{2004-April}
\title{Fractal structure of a white cauliflower}
\author{Sang-Hoon Kim}
\email{shkim@mail.mmu.ac.kr}
\affiliation{Division of Liberal Arts,
Mokpo National Maritime University, Mokpo 530-729, {\rm and}
 Institute for Condensed Matter Theory, Chonnam National University,
Gwangju 500-757, Korea}
%%%%%%%%%%%%%%%%%%%%%%%%%%%%%%%%%%%%%%%%%%
\date{\today}
%%%%%%%%%%%%%%%%%%%%%%%%%%%%%%%%%%%%%%%%%%%
\begin{abstract}
The fractal structure of a white cauliflower is investigated by
box-counting method of its cross-section. The capacity dimension
of the cross-section is $1.88 \pm 0.02$ independent of
directions. From the result, we predict that the capacity
dimension of the cauliflower is about 2.8. The vertical
cross-section of the cauliflower is modeled into a self-similar
set of a rectangular tree. We discuss the condition of the fractal
object in the tree, and show that the vertical cross-section has an
angle of 67 degrees in our model.
\end{abstract}
\maketitle
%\begin{multicols}{2}
%%%%%%%%%%%%%%%%%%%%%%%%%%%%%%%%%%%%%%
\section{Introduction}
%%%%%%%%%%%%%%%%%%%%%%%%%%%%%%%%%%%%%%%
A cauliflower is a variety of cabbage with an edible head of
condensed flowers and flower stems. it is a form of
cabbage in the mustard family, consisting of a compact terminal
mass of greatly thickened, modified, and partially developed
flower structures, together with their embracing fleshy stalks.
This terminal cluster has been known as a typical fractal among
living organisms\cite{man}.

It is clear that the dimensionality of a fractal is the most
fundamental concept of fractal analysis. Nevertheless, the fractal
dimension of a white cauliflower has not known yet. Grey and
Kjems discussed the fractal dimension of a cauliflower, and they
just suggested the possibility that the fractal dimension of a
cauliflower could be larger than that of a broccoli\cite{grey}.
Romera et. al. suggested a mathematical model of a cauliflower as
a sequence of a baby Mandelbrot set \cite{rom}, and its fractal
dimension was not obtained, either.

In this article, we measure the fractal dimension of the
cross-section of a white cauliflower. 
Then, the fractal dimension of the bulk cauliflower is introduced.
Next, we create a mathematical model for the cauliflower,
 and compare it with our experimental measurement.
The condition of creating the fractal
in our mathematical model is discussed, too.
 
%%%%%%%%%%%%%%%%%%%%%%%%%%%%%%%%%%%%%%
\section{fractal dimension of a cross-section}
%%%%%%%%%%%%%%%%%%%%%%%%%%%%%%%%%%%%%%%

There are many definitions of fractal dimensions, and the
basic concept in our model has almost the property as follows:
 Let us introduce that  $\delta$ is a measurement
scale, and that the measurement is $M_{\delta}(F)$, then the fractal
dimension $D$ of a set $F$ is determined by the power law 
\begin{equation}
M_{\delta}(F) \sim \delta^{-D}.
\label{10}
\end{equation}
If $D$ is a constant as $\delta \rightarrow 0$, 
$F$ has a dimension of $D$. 

The term ``capacity dimension" or
``box-counting dimension" is most widely used, 
because it can be easily applied to
fractal objects. It is defined as
\begin{equation}
D = \frac{\log N_{\delta}(F)}{-\log \delta},
\label{20}
\end{equation}
where $N_{\delta}(F)$ is the smallest number of sets of diameter
at most $\delta$ which can cover $F$\cite{fal1,fal2}.

Fractal objects embedded in one- or two-dimension are easy to
measure relatively compared to other higher dimensions, but most
of fractals in nature that are embedded in three dimensions 
is difficult to measure even in the capacity dimension. However,
 the dimensions of a cross-section is generally known to be related with
that of the bulk. 
Let us think of a bulk of three dimensions, and
assume the dimension of its cross-section is two, that is, it is
2/3 of the bulk. We can propose an  {\it ansaz} from
the above property: Let $D_c$ be a fractal dimension of a
cross-section embedded in two dimensions. Then, the fractal dimension 
of the bulk imbedded in three dimensions can be written as
\begin{equation}
D = \frac{3}{2N}\sum_{i=1}^{N} D_{ci},
\label{30}
\end{equation}
where $i$ represents every possible cross-section.
As the fractal dimension of the cross-section is independent of 
directions, Eq. (\ref{30}) is simply written as $D=(3/2)D_c$.

%%%%%%%%%%%%%%%%%%%%%%%%%%%%%
\section{experimental measurement}
%%%%%%%%%%%%%%%%%%%%%%%%%%%%%

We prepared several white cauliflowers of about $300g$ in mass 
and $15cm$ in diameter. We cut half of them in horizontal (H)
direction, and the other half in vertical (V) direction.
Then, we scanned the cross-section by a scanner 
and read the image into black and white in Fig. 1.
Here, the two figures by the two perpendicular
directions look totally different.

Next, we read the image into matrix of numbers.
The numbers of the matrix are the reflectivity of each pixels. 
The black backgrounds produces 0's, and 
the white images produce large numbers. 
Let us call the non-zero number 1 for convenience.
The size of the matrix is about 
$M_1 \times M_2 = 1400 \times 1200$. 
Note that the size of a pixel is an order of 100$\mu m$. 

In order to measure the fractal dimension of the cross-sections 
in Eq.\ (\ref{20}), we count the non-zero numbers 
by the box-counting method.
It becomes the smallest number of sets to cover the white images
by $100 \mu m \times 100 \mu m$.
In a half reduction procedure of the matrix, 
a $4 \times 4$ component becomes 1 
 if it contains any non-zero number, 
otherwise it becomes 0. 
Then, the number of 1's becomes the smallest number of
sets to cover the white images by $200 \mu m \times 200 \mu m$.
And so on.
In the reducing steps in half, that is
$M_1/2^n \times M_2/2^n, n=1,2,3...$,
the conversion from the reduced matrices to images
is plotted in Fig. 2 for the first four steps.
We see that the basic structure of the cross-sections
remains unchanged.

Fig. 3(a) is the log-log plot for the H-direction,
and Fig. 3(b) is for the V-direction.
We plotted it for a couple of different cauliflower samples.
The slopes are the capacity dimensions of the cross-sections.
Surprisingly, we found the two slopes for different directions are
very similar. It is $1.88 \sim 1.90$. Repeating the procedure for
different white cauliflowers, we observed similar values of
$D_{c}$ as $ 1.88 \pm 0.02$ for the two different directions.
Therefore, from Eq.\ (\ref{30}) we predict that the capacity
dimension of the white cauliflower is about 2.8.

%%%%%%%%%%%%%%%%%%%%%%%%%%%%%%%%%%%
\section{Mathematical modeling}
%%%%%%%%%%%%%%%%%%%%%%%%%%%%%%%%%%%

Converting a fractal found in nature into a corresponding
mathematical model is a hope of a theorist, but most of the work
is far away from the real world or extremely complicated in
analysis. 
The modeling of nature should be as simple as it can,
and at the same time, it should contain the basic structure of
nature. Furthermore, if some physical quantities of the model can
match the real systems, it will guarantee more credits to the
mathematical model.

We modeled the cross-section of a cauliflower in V-direction as a
rectangular tree that has three equilateral sides in Fig. 4.
 This is the simplest model of a self-similar set that has a single
scale factor $s$. 
Note that the cross-section of a broccoli in
V-direction has modeled into a self-similar set of triangular
tree or Pythagoras tree \cite{pei}.

The the scale factor $s$ of the rectangular tree
in Fig. 4 have the following relation
\begin{equation}
s=\frac{1}{2\cos \phi + 1},
\label{40}
\end{equation}
where $\phi$ is the angle of the rectangular tree.
Downsizing the tree by a ratio $s$, it creates three branches.
Therefore, the fractal dimension of Fig. 4 is
\begin{equation}
D_c =\frac{\log 3}{-\log s}.
\label{45}
\end{equation}

The areas of the rectangular tree have the following series
\begin{equation}
1, 3 s^2, 9 s^4, 27 s^6,... (3 s^2)^n,...
\label{50}
\end{equation}
where $n=0, 1, 2, 3, ...$. Because the fractal is 
the limit of the series, it should converge as $n$ increases. 
The condition of convergence of the series is 
$3 s^2 <  1$, or $s < 0.577$.  
It is clear from $D_c < 2$.
This condition corresponds to $\phi < 68.5^\circ$ 
by Eq.\ (\ref{40}). 

Since the $D_c$ in the cauliflower is 1.88, the scale factor $s$
and the angle $\phi$ in the model is obtained as 0.56 and
$67^\circ$ from Eqs.\ (\ref{40}) and (\ref{45}).
We recognize that the cross-section in
V-direction of the cauliflower is pretty close to two dimensions.

As the $\phi$ approaches $68.5^\circ$, the dimension of the tree
goes to two because the limit bents to cover the two dimensional
surface. On the other hand, as the $\phi$ approaches $0^\circ$,
the dimension of the tree goes to one because the limit goes to a
long rod. The head of the real cauliflower is rounded
not like the model in Fig. 4. It gives a possibility that the
real cauliflower is not a self-similar set of single scale factor,
but a set of multi-scale factors.

%%%%%%%%%%%%%%%%%%%%%%%%%%%%%%%%%%%%
\section{summary}
%%%%%%%%%%%%%%%%%%%%%%%%%%%%%%%%%%%%

We measured the fractal dimension of the cross-section of a white
cauliflower by the direct scanning method. 
It was $1.88 \pm 0.02$, and almost
independent of the directions of the cross-sections. From these
results, we predict that the fractal dimensions of the bulk
cauliflower is about 2.8.

We created a mathematical model for the V-direction of a
cauliflower with only one scale factor.
It is a rectangular tree of three equilateral sides.
We suggested the condition of creating fractals in our model,
and we showed that the angle of the model is 67 degrees
 and the scale factor is 0.56 comparing with experiment. 
This method of a scanning cross-sections 
and mathematical model from a polygonal tree
can be widely applied to many complex 
bulk fractals in nature. 
 
%%%%%%%%%%%%%%%%%%%%%%%%%%%%%%%%%%
\acknowledgements
%%%%%%%%%%%%%%%%%%%%%%%%%%%%%%%%%
We send our special thanks to K.S. Kim for useful discussions 
and  J. K. Ahn for graphic assistance.
%%%%%%%%%%%%%%%%%%%%%%%%%%%%%%%%%%%%%%
%%%%%%         references     %%%%%%%%%%%%%%%%%%%
%%%%%%%%%%%%%%%%%%%%%%%%%%%%%%%%%%%%%%

%%%%%%%%%%%%%%%%%%%%%%%%%%%%%%%%
\newpage

%%%%%%%%%%%%%%%%%%%%%%%%%%%%%%%%%%
%%%%%%%   FIGURE CAPTIONS   %%%%%%%%%%%%%%%%%%
%%%%%%%%%%%%%    1     %%%%%%%%%%%%%%%%
%\begin{figure*}
 \begin{figure}
%%%%%%%%%%%%%%%   2    %%%%%%%%%%%%%%%%%%%%%%%%%%%%%%%%%%%%
 \caption {The scanned images of the two cuts of the cauliflower. 
(a) Horizontal direction, (b) Vertical direction.  }
%%%%%%%%%%%       3     %%%%%%%%%%%%%%%%%%
 \caption { Converted images of the cross-section 
 by the half reduction procedure. 
The upper one is the horizontal direction (a), 
 and the bottom one is the vertical direction (b).  
%From the top left and clockwise direction,
(1) $M_1/2 \times M_2/2,$ (2) $ M_1/4 \times M_2/4$,
(3) $M_1/8 \times M_2/8,$ and (4) $ M_1/16 \times M_2/16.$ }
%%%%%%%%          4         %%%%%%%%%%%%%%%%%%%%
% \begin{figure}
 \caption { The log-log plots of the two perpendicular directions
for two different cauliflower samples.
 The slopes are $D_{c}'s$ of the white cauliflower. 
(a) Horizontal direction, (b) Vertical direction. }
%\end{figure}
%%%%%%%%%%%      5           %%%%%%%%%%%%%%%%
% \begin{figure}
\caption {Self-similar set of a rectangular tree 
with only one scale factor: $s= 0.56.$ or $ \phi=67^\circ$}
\end{figure}
%%%%%%%%%%%%%%%%%%%%%%%%%%%%%
%\end{multicols}
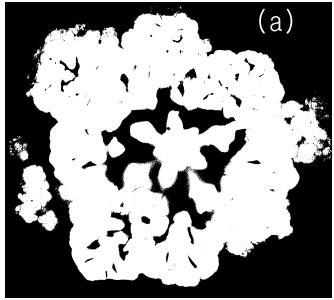
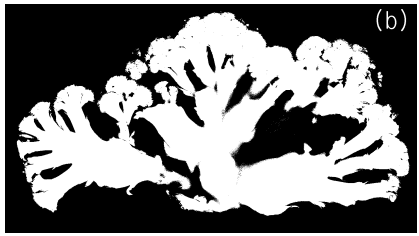
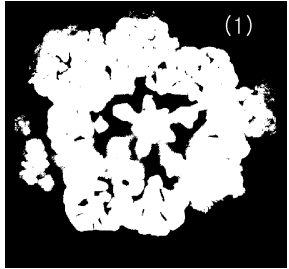
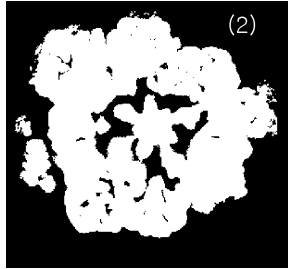
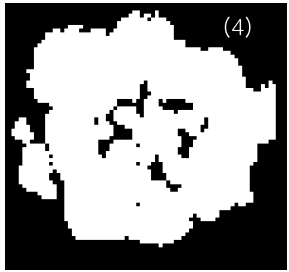
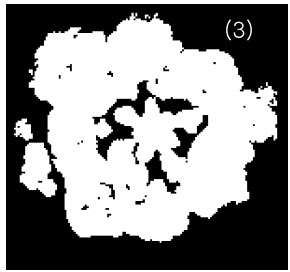
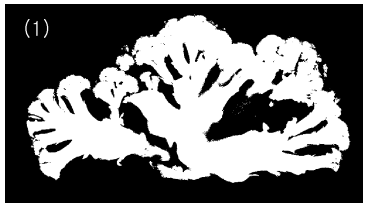
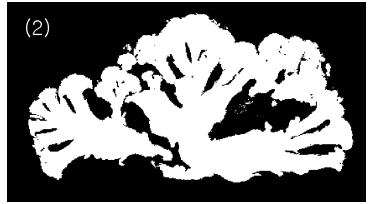
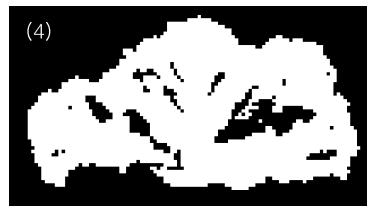
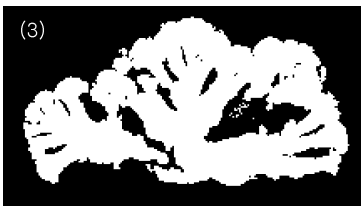
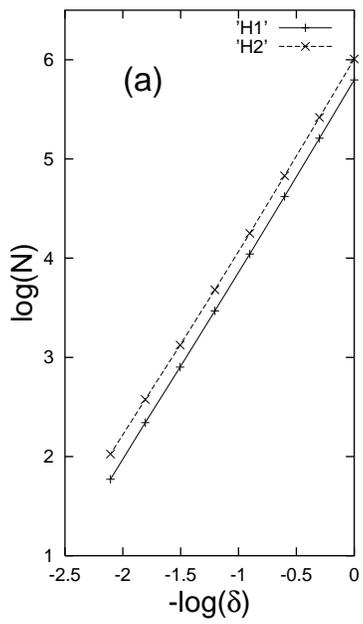
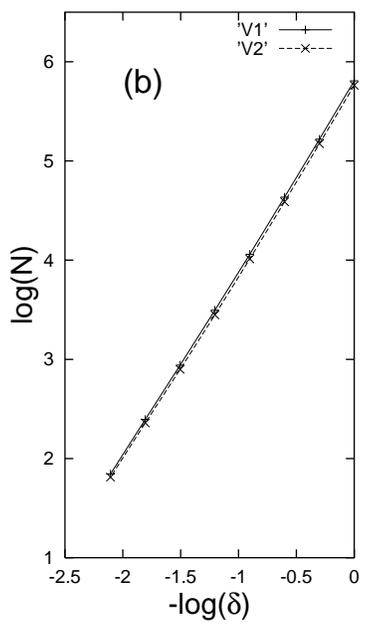
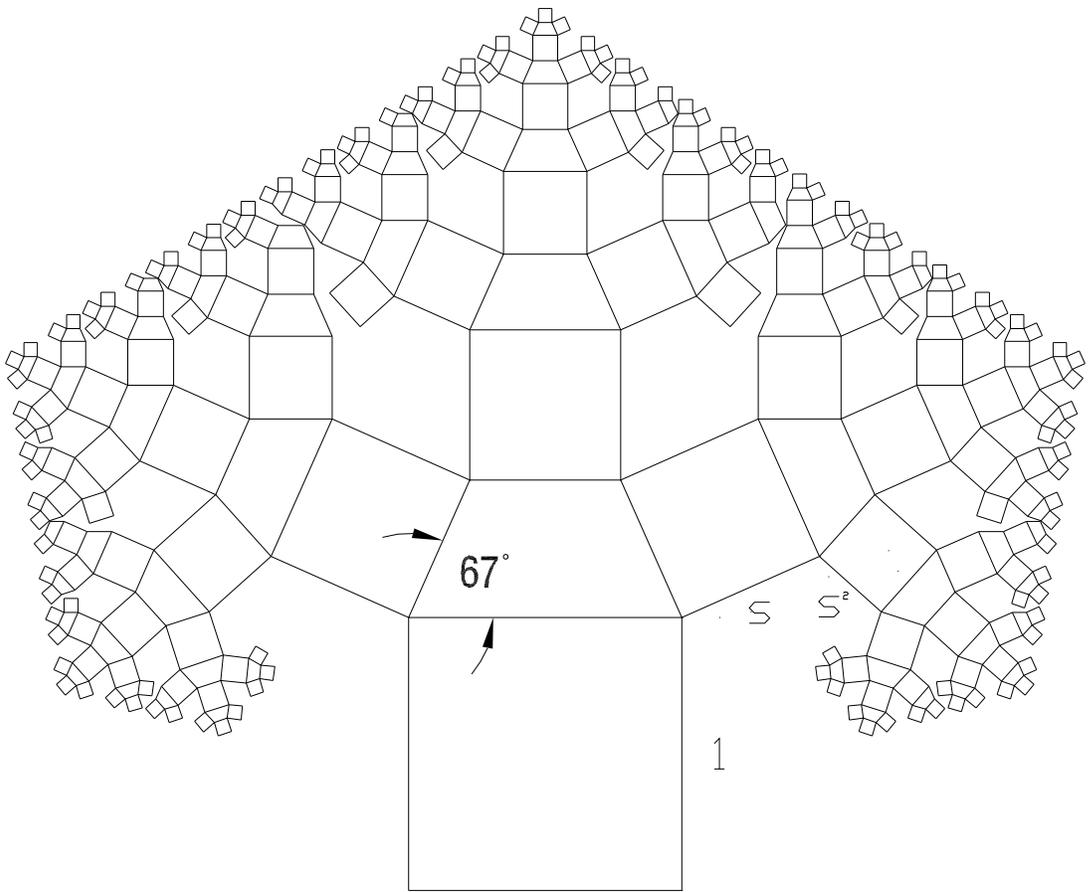
\end{document}